\documentclass[aps,pre,twocolumn]{revtex4}
\usepackage{bm}
\usepackage{color}
\usepackage{amsmath}
\usepackage{graphicx}
\usepackage{amssymb}

\begin{document}

\title{The chiral magnetic nanomotors}

\author{Konstantin I. Morozov}
\author{Alexander M. Leshansky}
\email{lisha@tx.technion.ac.il}
\affiliation{Department of Chemical Engineering, Technion -- Israel Institute of Technology, Haifa 32000, Israel}

\begin{abstract}

Propulsion of the chiral magnetic nanomotors powered by a rotating magnetic field is in the
focus of the modern biomedical applications. This technology relies on strong interaction
of dynamic and magnetic degrees of freedom of the system. Here we study in detail various experimentally observed regimes of the helical nanomotor
orientation and propulsion depending on the actuation frequency, and establish the relation of these two properties with the remanent magnetization and geometry of the helical nanomotors. The theoretical predictions for the transition between the regimes and nanomotor orientation and propulsion speed are in excellent agreement with available experimental data. The proposed theory offers a few simple guidelines towards the optimal design of the magnetic nanomotors. In particular, efficient nanomotors should be fabricated of hard magnetics, e.g., cobalt, magnetized transversally and have the geometry of a normal helix with a helical angle of $35^\circ\div 45^\circ$.

\end{abstract}

\maketitle

\section*{Introduction}

The development of micro- and nano-robots that can be remotely actuated, navigated
and delivered to a given location within the human body,
is a leading objective of the modern biomedical applications \cite{N1}.
The idea of using the external magnetic fields for in vitro and in vivo micromanipulation was recognized
long ago with the development of the ferrofluids, or colloidal solutions of magnetic
nanoparticles \cite{Rosen}. For a long time, however, the magnetic drag targeting
was achieved by the traditional technique based on applying the \emph{gradient}
magnetic field  \cite{Lubbe}. Despite some progress in this field, the method
proved to be not very effective and convenient primarily due to the need to operate with large gradients of the applied field.
The situation has changed dramatically with the development of
the fundamentally new approach  to the directed transport of magnetic particles.
It was demonstrated \cite{GF,N2,N3} that the rotating magnetic field even of small or moderate amplitude
can be applied for propulsion of \emph{chiral} magnetic nanoparticles.
These particles are named ``artificial bacterial flagella" due to their biomimetic helical shape that provides chirality necessary for
propulsion \cite{N2}.

The typical propulsion velocities offered by the new technique prove to be four-five orders of magnitude higher than
these achieved by the gradient methods (see scaling arguments in the Supporting Information). It is not surprising
that in the past few years the new technique has attracted considerable attention \cite{Tot,G,Peyer,G_new}.
Physically, the dynamics of the chiral magnetic nanomotors involve two interrelated phenomena of
magnetic and hydrodynamic nature, respectively.  It was found in experiments \cite{Tot,G}
that at low frequency of the rotating magnetic field the helices wobble
about the axis of the field rotation and their propulsion is hindered. However, upon increasing the field frequency, the wobbling
gradually diminishes and a corkscrew-like propulsive motion takes over.
Therefore a general search for the optimal nanomotor design should address
the questions of optimal choice of magnetic materials, magnetization procedure and helical geometry. In the recent investigation
\cite{Keaveny} one particular aspect of the problem was addressed, namely, a shape optimization of the swimmer geometry maximizing propulsion velocity at given applied magnetic torque assuming perfect alignment of the helix
along axis of the field rotation. In the present study we establish the deep interrelation
between the hydrodynamic and magnetic counterparts of the problem and show how it affects the dynamics of nanomotors.

\section*{Magnetized helix in rotating magnetic field: problem formulation}

\begin{figure}[h] \centering
\includegraphics[width=0.3\textwidth]{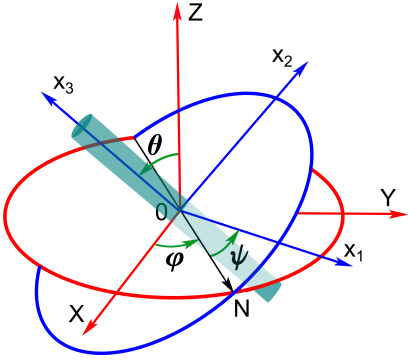}
\caption{
The laboratory and body-fixed coordinate systems with the corresponding axes
$XYZ$ and $x_1x_2x_3$ and the definition of the Euler
angles $\varphi$, $\theta$ and $\psi$. For simplicity, the cylindrical
envelope of a helix is shown.
\label{fig:1}}
\end{figure}

Lets consider a behavior of magnetized helix in a rotating
magnetic field. We assume that the longitudinal and transverse
components of a helix magnetization are fixed and equal
to $\mu$ and $m$, respectively. We will consider the problem in
two different coordinate systems -- in the laboratory coordinate system (LCS)
fixed in space and in the body-fixed coordinate system (BCS) rigidly attached
to the cylinder mimicking the helix. We study a helix behavior in the externally imposed
rotating uniform magnetic field $\bm{H}$. We assume that in the LCS
the field rotates in the horizontal plane
\begin{equation}
\bm{H}=H(\cos \omega t, \sin \omega t, 0) \,, \label{eq:field}
\end{equation}
where $H$ and $\omega$ are, correspondingly, the field amplitude and its frequency.
Specifying both coordinate systems, we will follow to the notation
given in the book \cite{LL}. The coordinate axes of the systems are
$XYZ$ and $x_1x_2x_3$, respectively. We choose $x_1$- and $x_3$ axes
of the BCS along the magnetization vectors $\bm{m}$ and ${\boldsymbol {\mu}}$, thus
the magnetic moment ${\boldsymbol {\mathcal {M}}}$ of the helix in the BCS has a form
\begin{equation}
  {\boldsymbol {\mathcal {M}}}^{BCS}=(m, 0, \mu) \,, \label{eq:moment}
\end{equation}
The orientation of the BCS relatively to the LCK is determined by the
Euler angles $\varphi$, $\theta$ and $\psi$ \cite{LL} shown in Fig.~\ref{fig:1}.

The magnetized helix is engaged in a rotational motion driven
by the magnetic torque $\bm{L}_m={\boldsymbol {\mathcal {M}}}\times\bm{H}$.
This torque is a source of both rotational and translational movements of the
particle. In the Stokes approximation, the helix motion is governed
by the balance of external and viscous forces and torques
acting on the particle \cite{HB}
\begin{eqnarray}
0= {\boldsymbol {\xi}}\cdot{\bm U}+ {\boldsymbol {\mathcal {B}}}\cdot \bm{\Omega}\,, \label{eq:u1}\\
\bm{L}_m={\boldsymbol {\mathcal {B}}}^T\cdot{\bm U}+ {\boldsymbol {\kappa}}\cdot\bm{\Omega}\,, \label{eq:u2}
\end{eqnarray}
where $\bm{U}$ and $\bm{\Omega}$ are the translational and angular velocities of helix, ${\boldsymbol {\xi}}$, ${\boldsymbol {\mathcal {B}}}$ and ${\boldsymbol {\kappa}}$ are the corresponding viscous friction tensorial coefficients. We have also assumed in Eq.~(\ref{eq:u1}) that there is no external force acting on the helix.

The formal solution of the problem can be readily obtained from Eqs.~(\ref{eq:u1}), (\ref{eq:u2}):
\begin{equation}
{\bm U}=-{\boldsymbol {\xi}}^{-1}\cdot{\boldsymbol {\mathcal {B}}}\cdot\bm{\Omega}\,,\,\,\,\bm{\Omega}={\boldsymbol {\kappa}_\mathrm{eff}^{-1}\cdot\bm{L}_m}\,, \label{eq:u3}
\end{equation}
where $ {\boldsymbol {\kappa}}_\mathrm{eff}= {\boldsymbol {\kappa}}- {\boldsymbol {\mathcal {B}}}^T \cdot{\boldsymbol {\xi}}^{-1}\cdot {\boldsymbol {\mathcal {B}}}$ is the renormalized matrix of the rotational friction.

Thus, the problem of the helix dynamics can  be divided into two separate problems: (i) rotational
motion of achiral slender particle (i.e. ${\boldsymbol {\mathcal {B}}}=0$), e.g. rod, and (ii) translation of a chiral particle rotating with a prescribed angular velocity.

In the following sections we consider both problems.

\section*{Magnetized cylinder in a rotating magnetic field}

It is convenient to right down the equation of the rotational
motion (the second equation in Eqs.~(\ref{eq:u3}))
in the body-fixed coordinate system in which the tensor ${\boldsymbol{\kappa}}$ is a
diagonal one. The components of the magnetic field (\ref{eq:field}) in the
BCS are  determined from the relation $ \bm{H}^{BCS}=\textbf{R}\cdot\bm{H}$,
where $\textbf{R}$ is the rotation matrix \cite{D} (see the Supporting Information).
Using data for the components of the angular velocity ${\bm{\Omega}}$ \cite{LL}
and elements of the rotation matrix, the torque balance equation
takes a form:
\begin{eqnarray}
Ac_{\varphi-\omega t}-Bs_{\varphi-\omega t}c_{\psi}s_{\theta}=\dot{\varphi}s_{\theta}\,, \label{eq:1}\\
As_{\varphi-\omega t}c_{\theta}+Bs_{\varphi-\omega t}s_{\psi}s_{\theta}=\dot{\theta}\,, \label{eq:2}\\
-C(c_{\varphi-\omega t}s_{\psi}+s_{\varphi-\omega t}c_{\psi}c_{\theta})=\dot{\varphi}c_{\theta}+\dot{\psi}\,. \label{eq:3}
\end{eqnarray}
Here $A=\mu H/\kappa_1$, $B=m H/\kappa_1$, and $C=m H/\kappa_3$ are the three characteristic
frequencies of the problem. We assume an arbitrary relation between the variables $\mu$ and $m$,
while $\kappa_3 < \kappa_1$, which is typical for a slender particle.
We also use the compact notation, i.e. $s_{\psi}=\sin{\psi}$, $c_{\theta}=\cos{\theta}$, etc. and the dot stands for the time derivative.

Generally, an overdamped dynamics of magnetic particles in a rotating magnetic field
can be realized via \emph{synchronous} and \emph{asynchronous} regimes \cite{Pincus,Cebers1}.
The synchronous regime is observed when there is a constant phase-lag between the Euler angle $\varphi$
of the particle body and the external magnetic field $\bm{H}$ while the angles $\theta$
and $\psi$ do not depend on time:
\begin{equation}
\varphi-\omega t:=\widetilde{\varphi}=\mathrm{const}\,,\,\psi=\mathrm{const}\,,\,\theta=\mathrm{const}\,. \label{eq:4}
\end{equation}
Despite the complex form the system (\ref{eq:1})-(\ref{eq:3}) admits a simple analytic solution in the synchronous regime. Indeed, substituting the ansatz (\ref{eq:4}) into  (\ref{eq:1})-(\ref{eq:3}) yields the following algebraic system
\begin{eqnarray}
Ac_{\widetilde{\varphi}}-Bs_{\widetilde{\varphi}}c_{\psi}s_{\theta}=\omega s_{\theta}\,, \label{eq:5}\\
s_{\widetilde{\varphi}}(Ac_{\theta}+Bs_{\psi}s_{\theta})=0\,, \label{eq:6}\\
-C(c_{\widetilde{\varphi}}s_{\psi}+s_{\widetilde{\varphi}}c_{\psi}c_{\theta})=\omega c_{\theta}\,. \label{eq:7}
\end{eqnarray}
Let us further consider the solutions of this system in detail.

\subsection*{Synchronous regime: low-frequency solution}

This obvious solution corresponds to the following values of the Euler angles: $\psi=0$,
$\theta=\pi/2$. For these values Eqs.~(\ref{eq:6}) and (\ref{eq:7}) are
satisfied identically, whereas Eq.~(\ref{eq:5}) reduces to a relation for the phase angle ${\widetilde{\varphi}}$:
\begin{equation}
Ac_{\widetilde{\varphi}}-Bs_{\widetilde{\varphi}}=\omega\,. \label{eq:8}
\end{equation}
The solution of the last equation is
\begin{equation}
{\widetilde{\varphi}}=\varphi_0- \arcsin (\omega/\omega_c^{(I)})\,, \label{eq:9}
\end{equation}
where $\omega_c^{(I)}=\sqrt{A^2+B^2}$ and $\varphi_0$ is the
angle between the total magnetic moment ${\boldsymbol {\mathcal {M}}}$  of the particle
and its transverse component $\bm{m}$ (see Fig.~\ref{fig:AB}): its cosine is given by
$c_{\varphi_0}=B/\sqrt{A^2+B^2}=m/\sqrt{m^2+\mu^2}$.

This solution corresponds to a rod rotation about its short axis
in the field plane, i.e., the angle $\theta$  between the long axis $x_3$ of the cylinder
and the axis $Z$ of the magnetic field rotation is equal to $\pi/2$ (see panel A in Fig.~\ref{fig:AB}). The solution is
physically clear. Indeed, in the low-frequency
regime the variations of the rod orientation
in space take place quasistatically: each instant the magnetic moment ${\boldsymbol {\mathcal {M}}}$ of
the rod follows the external magnetic field $\bm{H}$ with some slight phase-lag.
Thus, both vectors ${\boldsymbol {\mathcal {M}}}$ and $\bm{H}$ prove to be normal to the vector
${\boldsymbol{\omega}}$ of the field rotation.

This is the low-frequency solution of the problem. It belongs to the limiting interval of field frequencies $[0,\omega_c^{(I)}]$. When, $\omega=\omega_c^{(I)}$ the synchronous regime switches to the asynchronous one \cite{Cebers1}. The simplest way to understand this effect is to consider behavior of the angle $\alpha$ between the vectors ${\boldsymbol {\mathcal {M}}}$ and $\bm{H}$. Generally, the cosine of this angle reads
\begin{equation}
c_{\alpha}=\frac {{\boldsymbol {\mathcal {M}}}\cdot\bm{H}}{\mathcal{M}H}=c_{\varphi_0}(c_{\psi}c_{{\widetilde{\varphi}}}
-s_{\psi}s_{{\widetilde{\varphi}}}c_{\theta})+s_{{\varphi}_0}s_{{\widetilde{\varphi}}}s_{\theta}\,. \label{eq:angle}
\end{equation}

For the low-frequency solution with $\psi=0$, $\theta=\pi/2$ it reduces to
\begin{equation}
c_{\alpha}^{(I)}=c_{\varphi_0}c_{{\widetilde{\varphi}}}+s_{\varphi_0}s_{{\widetilde{\varphi}}}=c_{\varphi_0-{\widetilde{\varphi}}}=
\sqrt{1-\left(\frac {\omega}{\omega_c^{(I)}}\right)^2}\,, \label{eq:angle2}
\end{equation}
where we took into account Eq.~(\ref{eq:9}). Therefore, $\omega_c^{(I)}$ is the maximal
field frequency of the synchronized plane rotation of the rod.  When $\omega=\omega_c^{(I)}$,
the angle between vectors ${\boldsymbol {\mathcal {M}}}$ and $\bm{H}$ achieves its maximal value of $\pi/2$ and so the magnetic torque $\bm{L}_m$. A further increase of the field frequency leads to the breakdown of the synchronous rotation of the rod since the viscous forces can no longer balance the magnetic forces.

The solution we found, however, is not unique.
There is an additional solution that branches from the present one at the finite value of frequency $\omega=A$, i.e., prior to transition to the asynchronous regime. This solution will be called the high-frequency branch.

\subsection*{Synchronous regime: High-frequency solution}

The high-frequency solution is characterized by the Euler angle $\widetilde{\varphi}=0$. From the system (\ref{eq:5})-(\ref{eq:7}) one obtains \cite{note1}
\begin{equation}
s_{\theta}=\frac {A}{\omega}\,,\,\,s_{\psi}=-\frac {\omega}{C}c_{\theta}=-\frac {\sqrt{\omega^2-A^2}}{C}\,. \label{eq:sol2}
\end{equation}
The angle between the vectors ${\boldsymbol {\mathcal {M}}}$ and $\bm{H}$ reads
\begin{equation}
c_{\alpha}^{(II)}=c_{\varphi_0}c_{\psi}=\frac {B}{C} \frac {\omega_c^{(II)}}{\omega_c^{(I)}}
\sqrt{1-\left(\frac {\omega}{\omega_c^{(II)}}\right)^2}\,, \label{eq:angle3}
\end{equation}
where $\omega_c^{(II)}=\sqrt{A^2+C^2}$. Since we assumed, $\kappa_3 < \kappa_1$,
then $B<C$, and  $\omega_c^{(I)}<\omega_c^{(II)}$. Similarly to analysis above,  $\omega_c^{(II)}$ is the maximal frequency of
the synchronous rotation found for high-frequency solution.

\begin{figure}[h] \centering
\includegraphics[width=0.4\textwidth]{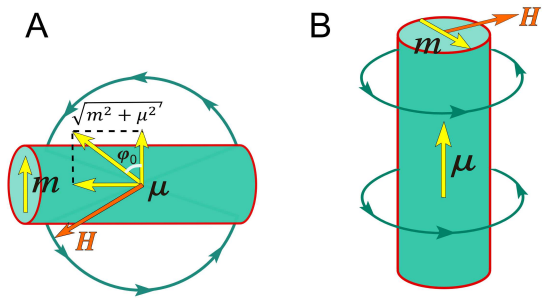}
\caption{
Two limiting geometries of the rod orientation relative to the axis of magnetic field rotation. A and B indicate low- and high-frequency regimes, respectively. To simplify the estimates
in high-frequency domain we assume $\theta \approx 0$, i.e., vertical orientation of the rod.
\label{fig:AB}}
\end{figure}

Which of these two synchronous solutions is selected at high frequencies?
To clarify this question qualitatively, let us consider two main limiting geometries shown
in Fig.~\ref{fig:AB}.
In the low-frequency regime, the horizontal plane of field rotation
coincides with the plane  formed by the magnetization vectors $\bm{m}$ and ${\boldsymbol {\mu}}$.
The vector of the total magnetic moment ${\boldsymbol {\mathcal {M}}}=\bm{m}+{\boldsymbol {\mu}}$ follows the external magnetic field $\bm{H}$
in attempt to instantaneously catch up with it and, thereby, \emph{minimizing the magnetic
energy} of the rotating rod. Taking into account Eqs.~(\ref{eq:angle}) and (\ref{eq:angle2}) the magnetic energy of the rod in the low-frequency regime reduces to
\begin{equation}
E_{magn}^{\,(I)}=-{{\boldsymbol {\mathcal {M}}}\cdot\bm{H}}=-\sqrt{(m^2+\mu^2)H^2-\omega^2 \kappa_1^2}\,. \label{eq:E1}
\end{equation}

In the high-frequency regime, the energy minimization is achieved differently:
the disadvantageous orientation of the longitudinal component ${\boldsymbol {\mu}}$
(${\boldsymbol {\mu}}\perp \bm{H}$) is compensated by the energy gain due to reduced friction associated with rotation about the long axis, compared to rotation about the short axis in the low-frequency regime,
\begin{equation}
E_{magn}^{\,(II)}=-{\bm{m}\cdot\bm{H}}=-\sqrt{m^2H^2-\omega^2 \kappa_3^2}\,. \label{eq:E2}
\end{equation}
Assuming for simplicity that $\kappa_1 \gg \kappa_3$, we estimate the minimal value of
the frequency $\omega \approx \mu H/\kappa_1=A$ starting from which the second regime becomes energetically favorable.

It should be emphasized however that in the above simplified analysis we assumed the orientation of the rod in the high-frequency regime to be vertical and ignored all the intermediate orientations. Actually, the high-frequency solution branches continuously from the low-frequency regime. Nevertheless, the above simplified analysis shows that the transition occurs at  $\omega = A$ in agreement with our estimate.

The aforementioned conclusion found from qualitative reasonings is confirmed also
by the rigorous stability analysis (see the Supporting Information).
It is interesting to note also that the lower values of dissipation correspond to
the high-frequency solution in comparison with that for the low-frequency one.

Thus, the following scenario of the phenomenon can be pictured. In the low-frequency regime, $0<\omega < A$, the rod keeps rotating synchronously in the plane of the applied magnetic field.
At higher frequencies, $A< \omega < \sqrt{A^2+C^2}$, the rod rotates also synchronously, while the procession angle $\theta$ between the long axis of the rod and the $Z$-axis of the field rotation diminishes with the increase in frequency. The minimal angle $\theta_\mathrm{min}$ is attained at $\omega=\omega_c^{(II)}$
\begin{equation}
(s_{\theta})_{min}=\frac {A}{\omega_c^{(II)}}=\frac {1}{\sqrt {1+\left(\frac {\kappa_1}{\kappa_3}\frac {m}{\mu } \right)^2}}\,. \label{eq:angle4}
\end{equation}

At $\omega=\omega_c^{(II)}$ the angle $\alpha$ between ${\boldsymbol {\mathcal {M}}}$ and $\bm{H}$
and the Euler angle $\psi$  attain their extremum $\alpha=-\psi=\pi/2$ and at even higher frequencies, $\omega>\omega_c^{(II)}$, the
high-frequency solution breaks down and the synchronous regime switches to the asynchronous one.

Finally, we note that the anticipated scenario of transitions between various regimes was observed for the first time in experiments \cite{Dhar} where the cylinder was magnetized only transversely, i.e., $\mu=0$. In the magnetic field rotating
horizontally the cylinder oriented vertically at high frequencies. The favorable at low frequencies horizontal orientation of the heavy rod was due to the action of gravity mimicking the longitudinal magnetization $\mu$ in our theory.

\subsection*{Asynchronous regime}

In the asynchronous regime all the Euler angles are time-dependent, ${\widetilde{\varphi}}={\widetilde{\varphi}}(t)$,
$\phi=\psi (t)$, $\theta = \theta (t)$, and the complete set of Eqs.~(\ref{eq:1})-(\ref{eq:3}) should be
considered. The system is complex and the analytical solution is not feasible. Meanwhile,
it admits a quite simple approximation. We shall present here a heuristic
approach to the solution of the problem and postpone the detailed analysis to future work.

The key element of our approach to the solution of the dynamical system (\ref{eq:1})-(\ref{eq:3})
at $\omega>\omega_c^{(II)}$ is the different character of time dependence of the three Euler
angles. It follows from Eqs.~(\ref{eq:1}) and (\ref{eq:2}) that the variables ${\widetilde{\varphi}}(t)$
and $\theta (t)$ are involved into oscillatory motion. The characteristic frequency $\Omega_{osc}$ of
oscillations is estimated easily at $\omega  \gtrsim \omega_c^{(II)}$
linearizing Eqs.~(\ref{eq:1}) and (\ref{eq:2}) around their limiting values (\ref{eq:sol2})
in the synchronous regime. It proves to be equal to $\Omega_{osc}=C\sqrt {1-B/C}\approx C$.
At the same time, Eq.~(\ref{eq:3}), determining the dependence $\psi(t)$ describes the appearance of the phase-lag of
particle rotation from the rotating magnetic field. The characteristic frequency $\Omega_{lag}$ of this
phase-lag is proportional to the supercriticality,  $\Omega_{lag} \sim C \sqrt{1-\omega_c^{(II)}/\omega}$,
(see below Eq.~(\ref{eq:c4})). Thus, at least in the vicinity of the critical frequency, $\Omega_{osc} \gg \Omega_{lag}$,
i.e., the angles ${\widetilde{\varphi}}(t)$ and $\theta (t)$ can be considered as `fast' variables, whereas $\psi(t)$ as a `slow'
one. Since we are interested only in the long-time dynamics, we can eliminate the fast variables by time-averaging of Eqs.~(\ref{eq:1})-(\ref{eq:3}). First two equations (\ref{eq:1}) and (\ref{eq:2}) reduce after coarsening again to Eqs.~(\ref{eq:5}) and (\ref{eq:6}) with the only difference that
instead of the dynamical variables ${\widetilde{\varphi}}(t)$ and $\theta (t)$ their time-averaged values $\overline{{\widetilde{\varphi}}}(t)$ and $\overline{\theta} (t)$ are used. The solution of these equations is similar to that considered in the previous section:
\begin{equation}
\overline{{\widetilde{\varphi}}}=0\,,\,\,s_{\overline{{\theta}}}=\frac {A}{\omega}\,. \label{eq:sola1}
\end{equation}

Using (\ref{eq:sola1}) in Eq.~(\ref{eq:3}) governing the dynamics of the slow variable $\psi(t)$ reduces to
\begin{equation}
\dot{\psi}=-\omega c_{\overline{{\theta}}}-Cs_{\psi}\,. \label{eq:c2}
\end{equation}
This is a standard form of the dynamical equation describing  plane rotation of a magnetic moment in rotating magnetic
field. It can be integrated analytically \cite{Pincus,Cebers1} to give
\begin{equation}
\psi(t)=-2\arctan \left[\gamma+\sqrt {1-\gamma^2} \tan \left(\textstyle \frac {\sqrt {1-\gamma^2}}{2}\omega c_{\overline{{\theta}}}t\right)\right]\,, \label{eq:c3}
\end{equation}
where $\gamma=C/(\omega c_{\overline{{\theta}}})=C/\sqrt {\omega ^2-A^2}$. The sign minus in the right hand side of
Eq.~(\ref{eq:c3}) means that the magnetic moment of particles lags behind the magnetic field ${\bm H}$.

The average value of the rotation frequency lag is obtained as $\Omega_{lag}=\overline{\dot{\psi}}=\lim_{T \rightarrow \infty}
(1/T)\int_0^T{\dot{\psi}}dt$ where $T$ is the period \cite{Cebers2}  and results in
\begin{equation}
\Omega_{lag}=-\sqrt {\omega ^2-{\omega_c^{(II)2}}}=-\sqrt {\omega^2-A^2-C^2} \,, \label{eq:c4}
\end{equation}

This completes the analysis of the rotational movement of the
magnetized rod. Let us next consider to the problem of the translational
motion of the helix actuated by the rotating external magnetic field.

\section*{Translational motion of the magnetized helix}

The translational motion (propulsion) of the rotating helix is due to its
chirality, i.e., to non-zero coupling resistance
matrix  ${\boldsymbol {\mathcal {B}}}$ (see Eq.~(\ref{eq:u3})). For simplicity we consider only the
helices with chirality along its axis, i.e., that in the BCS
the single non-zero coefficient of ${\boldsymbol {\mathcal {B}}}$ is $\mathcal{B}_{\|}$.
Therefore an $x_3$-projection of the translational velocity of helix is
$U_3^{BCS}=-(\mathcal{B}_{\|}/\xi_{\|})\Omega_3$.
In the LCS the velocity of the helix is
$ {\bm U}={\bf R}^T\cdot{\bm U}^{BCS}$,
where ${\bf R}^T$ is the transposed rotation matrix \cite{D}.
$\Omega_3$ is expressed in terms of the Euler angles as
$\Omega_3=\dot{\varphi}c_{\theta}+\dot{\psi}$ \cite{LL}.
Thus the translational velocity of helix in the rotating magnetic field is
\begin{equation}
{\bm U}=-(s_{\varphi}s_{\theta},-c_{\varphi}s_{\theta},c_{\theta})(\dot{{{\varphi}}}c_{\theta}+\dot{\psi})\mathcal{B}_{\|}/\xi_{\|} \,. \label{eq:vel}
\end{equation}

Let us consider the most relevant for propulsion $Z$-component of the velocity for regimes considered above.

\noindent \emph{Low-frequency synchronous regime.}

Since $\theta=\pi/2$ the propulsion velocity is zero by symmetry, $U_Z=0$.

\noindent \emph{High-frequency synchronous regime.}

From Eq.~(\ref{eq:sol2}) we obtain
\begin{equation}
U_Z=-{\omega}c_{\theta}^2\mathcal{B}_{\|}/\xi_{\|}=-{\omega}\left[1-({A}/{\omega})^2\right]\mathcal{B}_{\|}/\xi_{\|}\,. \label{eq:vel2}
\end{equation}

\noindent \emph{Asynchronous regime.}

From Eqs.~(\ref{eq:sola1}) and (\ref{eq:c4}) one finds:
\begin{equation}
U_Z=-{\omega}\left(1-\frac{A^2}{\omega ^2}\right)
\left(1-{\sqrt{1-\frac {C^2}{\omega ^2-A^2}}}\right)\mathcal{B}_{\|}/\xi_{\|}\,. \label{eq:vel3}
\end{equation}

\begin{figure}[h] \centering
\includegraphics*[width=0.2\textwidth]{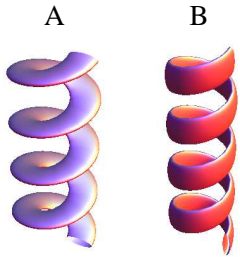}
\caption{
Illustration of the normal (A) and binormal (B) helical propellers.\label{fig:shapes}}
\end{figure}

\begin{figure}[h] \centering
\includegraphics[width=0.35\textwidth]{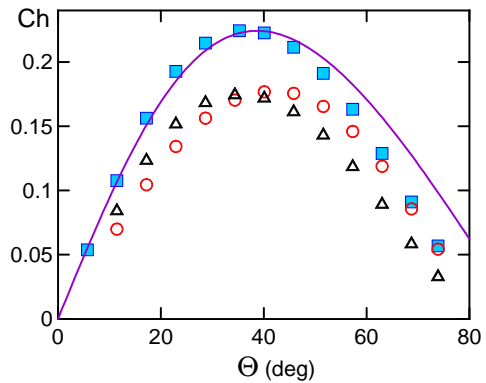}
\caption{
The chirality coefficient $\mathrm{Ch}=-\mathcal{B}_{\|}/(\xi_{\|}R)$ vs. the helical angle $\Theta$ for regular (circular cross-section, filled squares), normal (circles) and binormal (triangles) helices. The solid line stands for the RFT prediction.
\label{fig:U3}}
\end{figure}

The ratio of the corresponding resistance coefficients in (\ref{eq:vel2}--\ref{eq:vel3}), $\mathcal{B}_{\|}/\xi_{\|}$ can be determined, e.g., numerically.
We used the particle-based algorithm \cite{lesha09} to compute the resistance coefficients for three different types of helices. The helices differ by the orientation of their filament cross-section, assumed to be elliptical, relatively to the helix axis. For \emph{normal} and \emph{binormal} helices the filament cross-section is elongated, correspondingly,  perpendicularly and along the helical axis (see the Supporting Information for details). Geometric illustration of both these types of helices is provided in Figs.~\ref{fig:shapes} A and B, respectively. The degenerate case of \emph{regular} helices corresponds to filaments with circular cross-section. The dimensionless chirality coefficient $\mathrm{Ch}=-\mathcal{B}_{\|}/(\xi_{\|}R)$ is depicted in Fig.~\ref{fig:U3} vs. the helical angle $\Theta$ for regular (squares), normal (circles) and binormal (triangles) helices.

For all three types of helices their radius was set to $R/b=4$, where $b$ is the characteristic radius of the filament cross-sectional area \cite{note2}. The cross-sectional aspect ratio for normal and binormal helices is 1:2 and length-to-width ratio of all filament is $L/2b=30$.

The solid line stands for the prediction of the local Resistive Force Theory (RFT) corrected for the finite width of the filament. It can be readily seen that the regular helix is a better propeller than normal and binormal helices, however at large enough $\Theta\gtrsim 60^\circ$ the normal helix is as good as the regular one. The optimal helical angle maximizing the propulsion speed is in the range $\sim 35^\circ$--$45^\circ$. The agreement between the RFT prediction and the numerical results (for a regular helix) is excellent up to $\Theta\approx 40^\circ$. The details of the numerical algorithm and the RFT derivation are provided in the Supporting Information.

\section*{Comparison to experiments}

\begin{figure}[h] \centering
\includegraphics[width=0.35\textwidth]{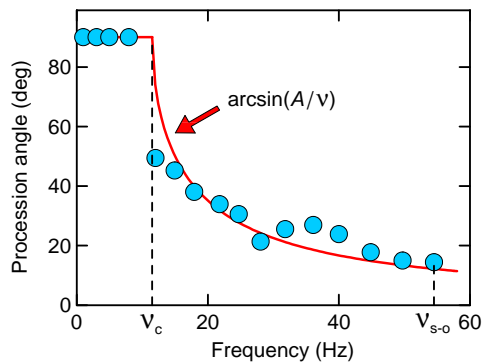}
\caption{
Procession angle as a function of frequency $\nu=\omega/2\pi$ of the rotating magnetic field. The experimental results (filled circles) are taken from \cite{G}. The solid line is the theoretical prediction in Eq.~(\ref{eq:sol2}). The dashed lines denote the critical frequency $\nu_c=11.5$~Hz and the step-out frequency $\nu_{s-o}=54.5$~Hz.
\label{fig:G1}}
\end{figure}

The comparison of the theoretical prediction for the procession angle $\theta$ in Eq.~(\ref{eq:sol2}) to the experimental results \cite{G}
is shown in Fig.~\ref{fig:G1} for the critical frequency $\nu_c=A=11.5$ Hz.
It can be readily seen that the experimental results are in an excellent agreement with the proposed theory.
The rightmost point in Fig.~\ref{fig:G1}
corresponds to the step-out frequency $\nu_{s-o}=\omega_c^{(II)}/2\pi=54.5$ Hz (marked by the dotted line)
at which the synchronous solution switches to the asynchronous one.

Let us estimate the limiting (minimal) value of the  procession angle, $\theta_{min}$,
corresponding to $\nu_{s-o}$. We approximate a helix by the prolate spheroid with the aspect ratio $a/b=4.5$ \cite{G}, where $a$ and $b$ stand for the major and minor semi-axis, respectively. The corresponding ratio of the rotational friction coefficients is $\kappa_1/\kappa_3=5.61$
(see the Supporting Information).
 The ratio $m/\mu$ of the components of magnetization is determined by the angle ${\varphi_0}$
(see Fig.~\ref{fig:AB}). According to \cite{G}, this angle equals to $\approx 54^{\circ}$, thus
$m/\mu=\cot 54^{\circ}=0.726$. Substituting both these values into Eq.~(\ref{eq:angle4})
yields $\theta_{min}=13.8^{\circ}$. The theoretical estimate proves to be close to the
the experimental value of $\theta_{min} \approx 14^\circ$ found in \cite{G} for the rightmost point in Fig.~\ref{fig:G1}.
A close estimate  $\theta_{min}=12.2^{\circ}$  can also be found from (\ref{eq:angle4})
using the frequency parameters $A=11.5$ Hz and $\nu_{s-o}$ mentioned above.

\begin{figure}[h] \centering
\includegraphics[width=0.35\textwidth]{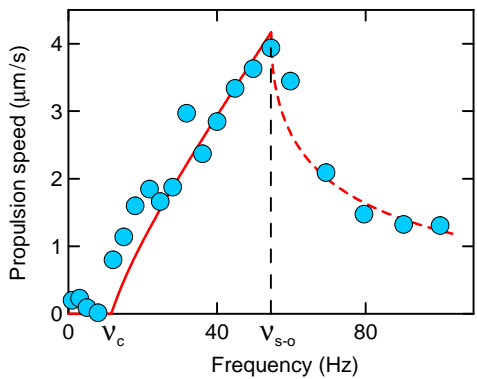}
\caption{ Propulsion velocity of a helix as a function of frequency $\nu=\omega/2\pi$ of the rotating magnetic field. The experimental results (filled circles) are taken from \cite{G}. The ascending (solid) line $f_1=0.08{\nu}(1-{\nu_c^2}/{\nu ^2})$ and descending (dashed) line $f_2=f_1\left(1-\sqrt{\nu^2-\nu_{s-o}^2}/\sqrt{\nu^2-\nu_c^2}\right)$ are the theoretical predictions from Eqs.~(\ref{eq:vel2}) and (\ref{eq:vel3}), respectively, with the parameters $\nu_c=11.5$~Hz and $\nu_{s-o}=54.5$ Hz. \label{fig:G2}}
\end{figure}

The comparison of the theory to the experimental results \cite{G} for the
the propulsion velocity $U_Z$ is shown in Fig.~\ref{fig:G2}.
Fitting the experimentally measured propulsion velocity in the high frequency synchronous regime (the solid line in Fig.~\ref{fig:G2}) to the theoretical prediction in Eq.~(\ref{eq:vel2}) we choose the value of the dimensional multiplicative coefficient $2\pi \mathcal{B}_{\|}/\xi_{\|}$ to be equal to $0.08$ $\mu$m. We determine the dimensionless ratio $\mathcal{B}_{\|}/(\xi_{\|}R)$ numerically as described in the Supporting Information and using the data reported in \cite{G}, i.e., $R=0.20$ $\mu$m the radius of the helix, $a=0.17$~$\mu$m for the radius of the filament, $P=1$~$\mu$m for the helical pitch, so that the helical angle $\Theta=\tan^{-1}(2\pi R/P)\approx 51.5^\circ$. The computed value of the dimensional coefficient $2\pi \mathcal{B}_{\|}/\xi_{\|}$ is equal to $0.094$~$\mu$m, while for a slightly higher angle $\Theta=53.5^\circ$ we find $2\pi \mathcal{B}_{\|}/\xi_{\|}=0.081$~$\mu$m that perfectly fits the experimental results. It is worth to also mention that the prediction for the propulsion velocity in the asynchronous regime in Eq.~(\ref{eq:vel3}) (the descending dashed line in Fig.~\ref{fig:G2}) does not involve any adjustable parameters and fits very well the experimental data.

\section*{Magnetization of helices}

In the preceding sections we studied the  movement of magnetized rod and/or helix in rotating magnetic field provided that the magnetization components $\bm m$ and $\bm\mu$ are prescribed.
In the present section we shall consider the general magnetic properties of the ferromagnetic helices.

The magnetism of helices is due to a thin layer ($ \lesssim 10$ nm) of ferromagnetic material (Ni, Co, Fe or their derivatives and alloys) coating a magnetically neutral helical body.
The magnetic properties of thin magnetic films are well studied for the case of \emph{plane}
films \cite{Good,Soohoo}. The shape of a thin film constrains the magnetization vector $\bm{M}$ (the magnetic moment
per unit volume) to lie in  the plane of the film \cite{Good}. The films are found to be \emph{uniaxially} anisotropic
ones, i.e., the magnetization vector $\bm{M}$ aligns with an \emph{easy} axis $\bm{e}$ in the absence of
the external magnetic field $\bm{H}$. The density $f_a=E_a/V$
of the anisotropy energy is written in the form:
\begin{equation}
f_a=K \sin ^2 (\widehat{\bm{M},\bm{e}})\,, \label{eq:ani}
\end{equation}
where $(\widehat{\bm{M},\bm{e}})$ is an angle between the vectors $\bm{M}$ and $\bm{e}$
and $K$ is an anisotropy constant. The absolute value $M$ of the magnetization is known -- it is equal
to its saturation value at given temperature. We indicate these values for three major magnetics at room
temperature: $M_{\mathrm {Ni}}=480$ G, $M_{\mathrm {Fe}}=1700$ G, $M_{\mathrm {Co}}=1400$ G.

The appearance of this in-plane anisotropy depends strongly on the film preparation conditions.
There are three possible factors affecting anisotropy: (1) short-range directional order of magnetic atoms during their deposition; (2) angle of incidence  between the substrate and the depositing
beam, and (3) anisotropic strain and other properties of substrate \cite{Good,Smith}. This is why the anisotropy is called \emph{induced} one vs. the uniaxial magneto-crystalline \emph{bulk} anisotropy. In particular, the induced magnetic anisotropy typically proves to be order of magnitude lower than its bulk value. Let us indicate the values of $K$ for three most common magnetics:  $K_{\mathrm {Ni}}=(1\div3) \cdot 10^3$ erg/cm$^3$ and $K_{\mathrm {Ni}}^{bulk}=5\cdot 10^4$ erg/cm$^3$ for nickel \cite{Soohoo}; $K_{\mathrm {Fe}}=3 \cdot 10^4$ erg/cm$^3$ and $K_{\mathrm {Fe}}^{bulk}=4.5\cdot 10^5$ erg/cm$^3$ for iron \cite{Knorr};  $K_{\mathrm {Co}}=4 \cdot 10^5$ erg/cm$^3$ and $K_{\mathrm {Co}}^{bulk}=4\cdot 10^6$ erg/cm$^3$ for cobalt \cite{Tanaka}.

The minimization of the sum $f_a+f_m$ of the anisotropy energy and magnetic energy $f_m=-\bm{M}\cdot\bm{H}$
determines the equilibrium orientation of vector $\bm{M}$ in the
combined field resulting from the
external magnetic magnetic field $\bm{H}$ and  the anisotropy field $\bm{H}_a=(2K/M)\bm{e}$ \cite{Good}.
The anisotropy field  $H_a$ defines the characteristic value
of external magnetic field required to film re-magnetizing. The characteristic values of the anisotropy field are $H_{a}^{\mathrm {Ni}}\sim 10$ Oe, $H_a^{\mathrm {Fe}}\sim 30$ Oe, $H_a^{\mathrm {Co}}\sim 500$ Oe. The  measured values of the anisotropy fields are mainly in agreement with the above estimate, however, they can scatter notably depending on the way of sample preparation. For example, $H_a$ of permalloy (Fe-Ni alloy) can vary from 1 to 50 Oe \cite{Smith}. Analogously, in the recent study \cite{Vergara} the in-plane uniaxial magnetic
anisotropy field of a multilayer cobalt film was found to be in the range from 150 to 900 Oe.

As we already mentioned, these results hold for the plane magnetic films. To the best of our knowledge there is no data for the ferromagnetic films of helicoidal geometry. Thus, aiming to estimate the \emph{remanent magnetization} of
magnetic helixes, we have to use the data obtained for the plane films under some additional assumptions.
We assume that (a) the magnetic layer covering the helix has a constant thickness; (b) the helix is composed
of integer number of turns; (c) the cross-section of helical filament is asymmetric.
For simplicity, we suppose also that the magnetic structure of the film is formed by an ensemble of magnetic domains with
the easy magnetization axes $\bm{e}$ determined by the helix geometry. We consider two potential orientations of the easy axis: (i)  $\bm{e}$ is the vector tangent to the centreline of the helix; (ii) $\bm{e}$ is directed along the longer cross-sectional axis of the helix. For the later case we have to consider two different geometries -- normal and binormal helices.
The details of calculations are given in the Supporting Information. Here we only provide the final results for the different particular cases.

\begin{figure}[h] \centering
\includegraphics*[width=0.35\textwidth]{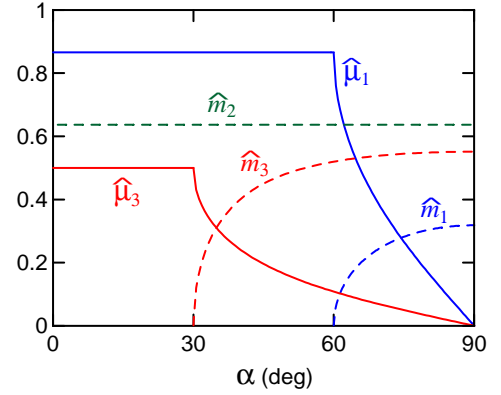}
\caption{
The dimensionless longitudinal $\widehat{\mu}$ (solid lines) and transversal $\widehat{m}$ (dashed lines) components of the remanent magnetization as a function of the angle between the magnetizing field and the helix axis.
Indexes correspond to three cases considered in the text.
The results are shown for the helix angle $\Theta=30^\circ$. In the case of the `dual' helix with the angle $\Theta=60^\circ$, one finds the `dual' solution: curves 1 and 3 interchange. For the helix with $\Theta=45^\circ$ curves 1 and 3 merge. Curve 2 remains unchanged for any value of $\Theta$.
\label{fig:mag}}
\end{figure}

\paragraph*{Easy axis is tangent to the helix centreline.}
We describe here the result for the longitudinal $\mu$ and transversal $m$ components of the remanent magnetization
appeared after removal of external field $\bm{H}=H(\sin \alpha, 0, \sin \alpha)$ which forms the angle $\alpha$
with the axis of helix. Both values are given in the units of the magnetic moment $M_sv$  of a hypothetical plane layer
with the same volume $v$ of magnetic material, $M_s$ is its saturation magnetization. The final result depends on two
angles -- the helix angle $\Theta$ and the magnetizing angle $\alpha$. For low values $\alpha$ of the angle
between the field $\bm{H}$ and the helix, $0<\alpha<\pi/2-\Theta$, the transversal component is absent:
\begin{equation}
\widehat{\mu}=\cos \Theta\,,\,\,\widehat{m}=0\,. \label{eq:m1}
\end{equation}
For high values $\alpha>\pi/2-\Theta$ there are both components:
\begin{equation}
\widehat{\mu}=\frac {2}{\pi}\cos \Theta \arcsin P\,,\,\,\widehat{m}=\frac {2}{\pi}\sin \Theta \sqrt {1-P^2}\,,
\label{eq:m2}
\end{equation}
where $P=\cot \alpha \cot \Theta$. The result is shown in Fig.~\ref{fig:mag}. We note
that Eq.~(\ref{eq:m2}) holds for both normal and binormal  helices.

\paragraph*{Easy axis is along the longer cross-sectional axis of the normal helix.}
This case of the normal helix with easy axis in the direction of the the longer cross-sectional axis is the
simplest one. The result does not depend on the orientation $\alpha$ of magnetizing field, there is only single
(transversal) nontrivial component of magnetization:
\begin{equation}
\widehat{\mu}=0\,,\,\,\widehat{m}=2/\pi\,. \label{eq:m3}
\end{equation}

\paragraph*{Easy axis is along the longer cross-sectional axis of the binormal helix.}

There are also the cases corresponding to small and large values of angle $\alpha$, respectively. For
small angles $\alpha < \Theta$ again there exists only the longitudinal magnetization:
\begin{equation}
\widehat{\mu}=\sin \Theta\,,\,\,\widehat{m}=0\,. \label{eq:m4}
\end{equation}
For high values of $\alpha>\Theta$ there are both components
\begin{equation}
\widehat{\mu}=\frac {2}{\pi}\sin \Theta \arcsin G\,,\,\,\widehat{m}=\frac {2}{\pi}\cos\Theta \sqrt {1-G^2}\,,
\label{eq:m5}
\end{equation}
where $G=\cot \alpha \tan \Theta$. We note that results for cases 1 and 3 are dual to each other, namely, Eqs.~(\ref{eq:m1}), (\ref{eq:m2})  found for some value
$\Theta=\Theta^*$ of the helix angle are identical to Eqs.~(\ref{eq:m4}), (\ref{eq:m5}) corresponding to the dual helix with the helix angle $\Theta_{dual}=\pi/2-\Theta^*$.

We consider here the  idealized case of prescribed orientation of
the easy axes of magnetic domains. In practice, there is a distribution of the easy axes in the plane film and
therefore a combined result is found for the limiting cases.
So, the magnetizing of normal helix proves to be a weighted result of cases 1 and 2. For binormal helix,
it is a combination of cases 1 and 3.

\section*{Concluding remarks}

To conclude, let us formulate a few simple rules for the optimal design of chiral magnetic nanomotors.
An effective steering of nanomotors should be obtained when they possess the maximal
transverse magnetization $m$ and minimal longitudinal magnetization $\mu$. The vanishing value of
$\mu$ allows to completely avoid wobbling and maximize the propulsion velocity
of nanomotors, $U_{max}\sim (mH/\eta V)R\:\mathrm{Ch}$, with $V$ being the propeller's volume. Since $m\sim M_s v$, the best propellers
are those (i) composed of the hard magnetics, e.g., cobalt with high value of the anisotropy field $H_a$; (ii) having high value of the saturation magnetization $M_s$; (iii)  magnetized transversally. The preferable geometry of helical nanomotors is probably
the normal helix (iv): the slightly lower values of the chirality $\mathrm{Ch}$ in comparison
with that of the regular helix (see Fig.~\ref{fig:U3}) is well compensated by the larger values of the magnetization as follows from Fig.~\ref{fig:mag}. The optimal value of the helix angle
is in the range $\Theta=35^\circ \div 45^\circ$ maximizing the chirality, $\mathrm{Ch}\approx 0.2$ (v).
Despite a large variability in the nanofabrication techniques and experimental setups \cite{GF,N2,N3,Tot,G,Peyer,G_new,Dhar}, it is interesting to point out that one of the pioneering experimental works in this field, \cite{GF}, has empirically adopted most of the rules formulated above.

\section*{Acknowledgement}

The authors would like to thank Ambarish Ghosh, Li Zhang and
Soichiro Tottori for providing details of the experiments with chiral
magnetic nanomotors and useful discussions. This work was partially supported by the Japan Technion Society
Research Fund (A.M.L.) and by the Israel Ministry for Immigrant Absorption (K.I.M.).


\begin{thebibliography}{99}
\bibitem{note1} In process of writing the paper, we became aware of the
article \cite{G_new} where the solution for sinchronous regime has also been found.
\bibitem{note2}
For normal and binormal helices $b$ is the length of the cross-sectional minor semi-axis.

\bibliographystyle{rsc} 
\end{thebibliography}
\end{document}